\pdfoutput=1
\documentclass{JINST}
\usepackage[latin1]{inputenc}
\usepackage{enumerate}
\usepackage{upgreek}
\usepackage{lineno}
\usepackage{amssymb}
\usepackage{array}
\usepackage{amsmath}

\title{Performance tests of an AGIPD 0.4 assembly at the beamline P10 of PETRA III}

\author{J. Becker$^a$\thanks{Corresponding author. E-mail: Julian.Becker@desy.de}~, A. Marras$^a$, A. Klyuev$^a$, F. Westermeier$^a$, U. Trunk$^a$, and H. Graafsma$^{a,b}$ \\ 
\llap{$^a$}Deutsches Elektronen-Synchrotron (DESY),\\
Notkestr. 85, 22607 Hamburg, Germany\\
\llap{$^b$}Mid Sweden University,\\
Holmgatan 10, S-85170 Sundsvall, Sweden\\
}

\abstract{
The Adaptive Gain Integrating Pixel Detector (AGIPD) is a novel detector system, currently under development by a collaboration of DESY, the Paul Scherrer Institute in Switzerland, the University of Hamburg and the University of Bonn, and is primarily designed for use at the European XFEL. To verify key features of this detector, an AGIPD 0.4 test chip assembly was tested at the P10 beamline of the PETRA III synchrotron at DESY. The test chip successfully imaged both the direct synchrotron beam and single 7.05 keV photons at the same time, demonstrating the large dynamic range required for XFEL experiments. X-ray scattering measurements from a test sample agree with standard measurements and show the chip's capability of observing dynamics at the microsecond time scale.
}


\begin{document}

\linenumbers

\section{Introduction}

The Adaptive Gain Integrating Pixel Detector (AGIPD) is a novel detector system currently under development by a collaboration of DESY, the Paul Scherrer Institute in Switzerland, the University of Hamburg and the University of Bonn \cite{AGIPD1, AGIPD2}. It was designed for use at the European XFEL \cite{XFEL, TN2011-001} and will be able to cope with the demanding requirements of this machine for 2D imaging systems. The European XFEL will provide pulse trains, consisting of up to 2700 pulses separated by 220~ns (600~$\upmu$s in total) followed by an idle time of 99.4~ms, resulting in a supercycle of 10~Hz and 27000 pulses per second. The energy of the X-rays will be tunable, and the tuning range will depend on the experimental station.

AGIPD is based on the hybrid pixel technology and aimed at imaging in the energy range between 3 and 15~keV. The current design goals of the newly developed Application Specific Integrated Circuit (ASIC) with independent dynamic gain switching amplifier in each pixel are a dynamic range of more than 10$^4$ 12.4~keV photons (for each pixel) in the lowest gain, single photon sensitivity in the highest gain, and operation at 4.5~MHz frame rate. 

Due to the special pulse structure of the European XFEL, it is necessary to store the acquired images inside the pixel circuit area during the pulse train. A compromise had to be found between storing many images, requiring a large pixel area, and high spatial resolution, requiring small pixel sizes \cite{AGIPD1, AGIPD2}. The AGIPD will feature a pixel size of (200~$\upmu$m)$^2$, which is sufficient to accommodate an analog memory for 352 images. An external veto signal can be provided to maximize the number of useful images by overwriting any image previously recorded during the pulse train. The image data is read out and digitized in the 99.4~ms between pulse trains.

Testing and calibration of novel detector systems is of essential importance for their performance. As it is very hard to approximate the properties of free electron laser sources using commonly available lab sources, beam tests at synchrotron sources are helpful to verify important functional parameters of the detector system. In the following we will present the results of a performance tests of AGIPD 0.4 at the synchrotron beamline P10 at PETRA III.

\section{The AGIPD 0.4 assembly}

\begin{figure}
	\centering
		\includegraphics[width=1.0\textwidth]{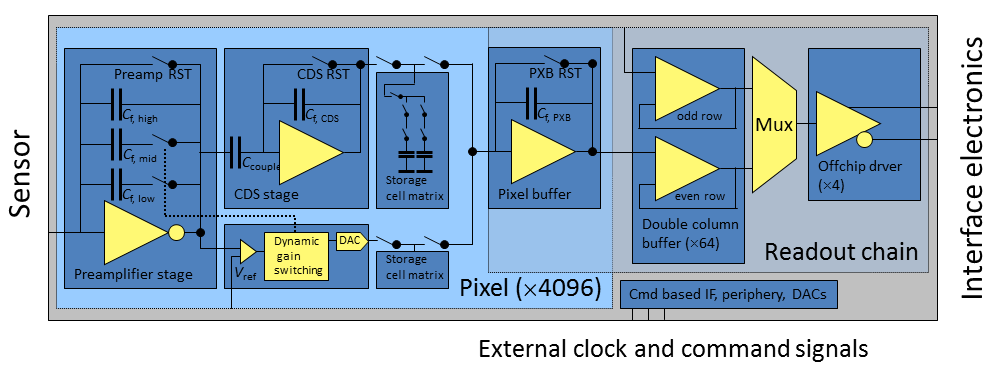} 
	\caption{Circuit diagram of AGIPD 1.0.}
	\label{ASIC}
\end{figure}

The AGIPD 0.4 assembly is a 16x16 pixel prototype of the AGIPD, bump bonded to a silicon sensor. The pixels have the same size as the final system, (200~$\upmu$m)$^2$, and feature the adaptive gain switching amplifier with 3 stages and 352 storage cells per pixel. A block diagram of the full scale chip is shown in Figure \ref{ASIC}. The detector assembly was mounted on a daughter card on the movable detector table and connected to a custom made chip tester box with cables\footnote{In this respect the setup is different from the usual lab setup. The additional noise introduced by the cables is minimal.}. The chip tester box handled all communication with the ASIC. In contrast to the sensor thickness intended for the full scale detector (500~$\upmu$m), the ASIC was bump bonded to a silicon sensor of 320~$\upmu$m thickness having a depletion voltage of  approximately 50~V. To (over-)deplete the sensor, a bias voltage of 120~V was applied. All measurements were performed without cooling at room temperature.

\begin{figure}
	\centering
	\begin{tabular}{@{}cc@{}} 
		\begin{tabular}{@{}c@{}} \includegraphics[width=0.45\textwidth]{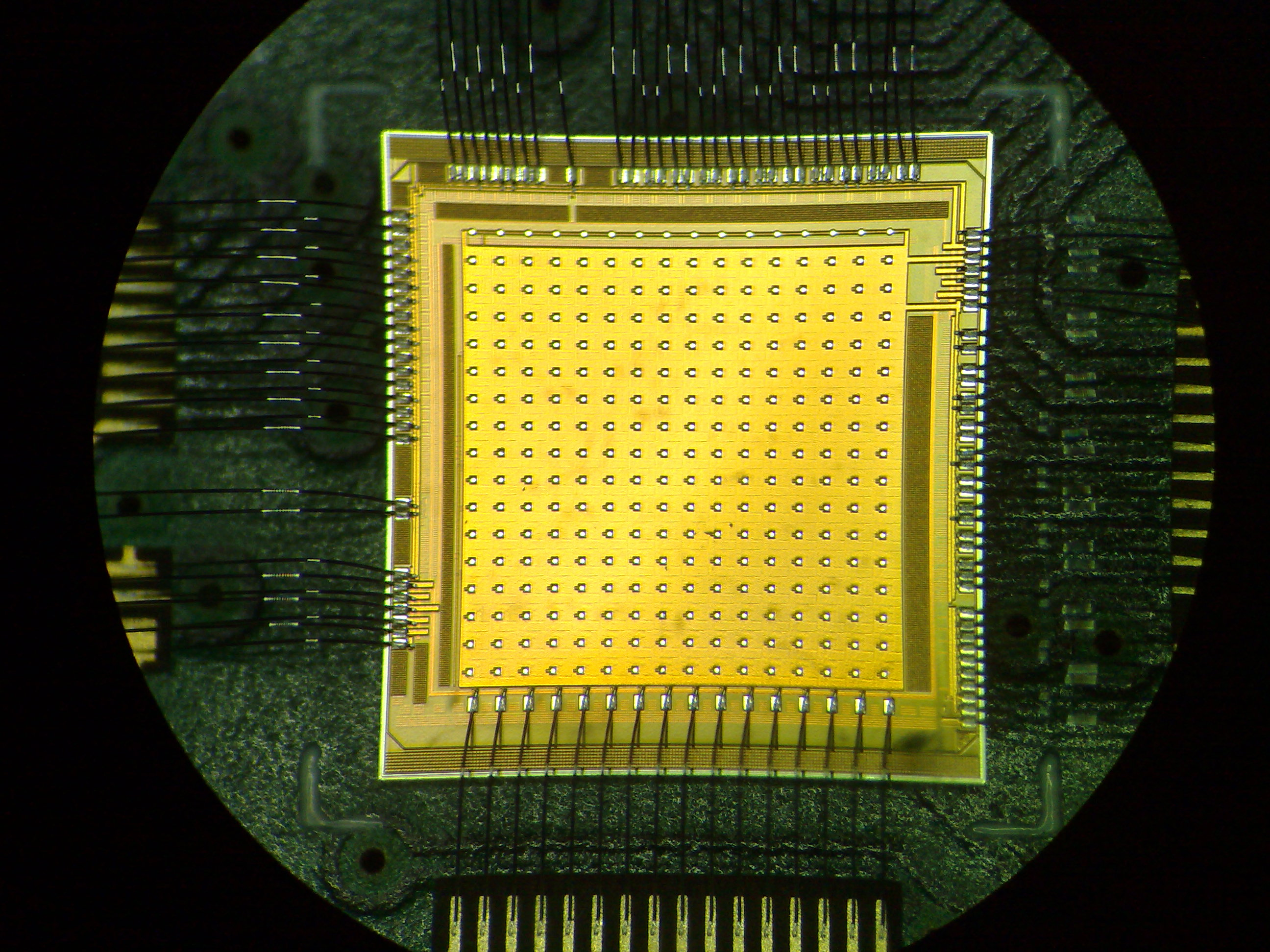}  \end{tabular} &
		\begin{tabular}{@{}c@{}} \includegraphics[width=0.45\textwidth]{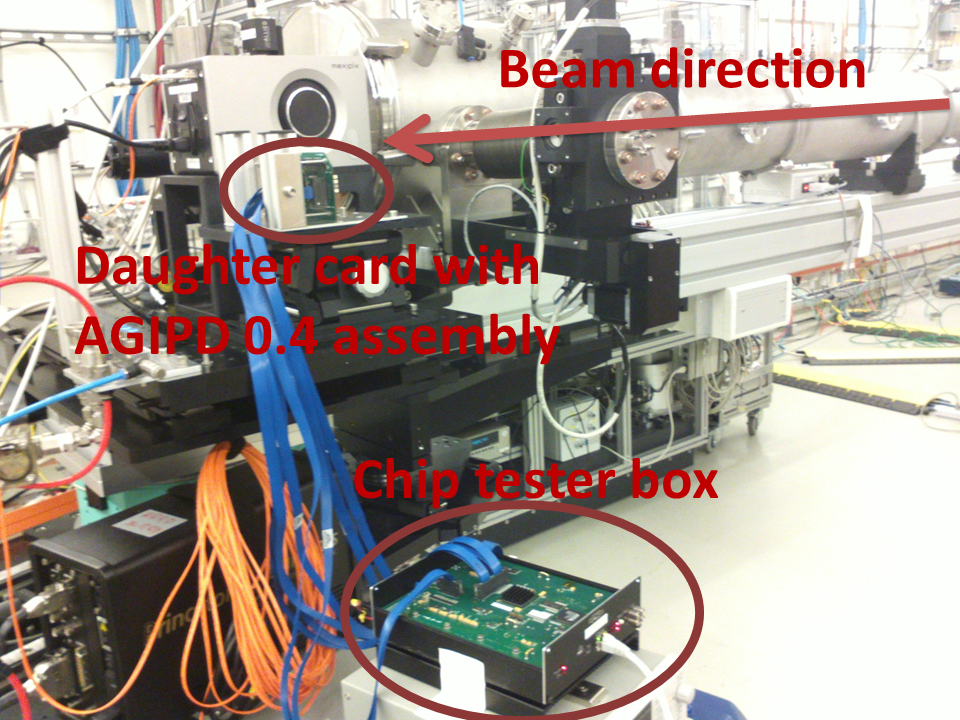}  \end{tabular} 
	\end{tabular}
	\caption{Left: Microscope picture of an AGIPD 0.4 chip without sensor mounted on a daughter card. Right: Annotated photograph of the experimental setup at the P10 beamline.}
	\label{photos}
\end{figure}

To visualize the chip and chip tester box a microscope picture and a photograph of the experimental setup at the beamline are shown in Figure \ref{photos}.



\subsection{Operation parameters and data acquisition}


As the chip tester box was designed for testing, communication and data acquisition is based on patterns (explained below) rather than images. Communication with the chip tester box occurs over a TCP/IP network.

The chip tester box communicates with the test chip by providing input signals (called vectors) on all input lines and digitizing the output of the analog output of the ASIC. All commands to the ASIC are constructed by sending a sequence of vectors. 

Some commands (e.g. the selection of a memory cell) require a fixed number of clock cycles to perform. Other commands, e.g. the reset of the preamplifier, require a defined time to perform and the number of required clock cycles depends on the master clock frequency.

\subsubsection{Patterns}

The vectors to be sent to the ASIC are generated inside the Field Programmable Gate Array (FPGA) of the chip tester box. This FPGA is programmed to send a pattern (a stream of vectors) to the ASIC. Due to the limited amount of memory inside the FPGA the maximum number of vectors in a pattern is limited. During the sending of the pattern the Analog-to-Digital Converter (ADC) on the chip tester board continuously samples the output of the ASIC into a FIFO (First In, First Out) buffer. Upon completion of the pattern the digitized output data is sent to the controlling PC over the network connection. A new acquisition cycle cannot start before the last data packet has successfully arrived at the PC.

Since the amount of commands sent to the ASIC in sequence is limited, the chosen master clock frequency of 5~MHz is a compromise between a high value, allowing high frame rates, and the number of commands that can be stored into the memory of the FPGA. It allowed to compose the following two patterns that were used for the experiments:

\begin{itemize}
\item Imaging pattern: This pattern was used for the experiments in imaging mode. It integrates for 200~ns (1 clock cycle) in all pixels before reading the analog and gain information of all 16x16 pixels in a defined sequence. Using this pattern frames could be written to disk with an average speed of 25 frames per second.
\item XPCS pattern: This pattern was used for the X-ray Photon Correlation Spectroscopy (XPCS) experiments. It integrates for 200~ns, then reads the analog information of a single pixel. This process is repeated 314 times, and the time between the beginning of two integration periods is 9.6~$\upmu$s.
\end{itemize}



Both patterns utilize only a single storage cell per pixel, which is immediately read after being written.

The chip can image at 4.5~MHz and more (using more than one storage cell per pixel), but as XPCS experiments critically depend on the number of frames recorded in sequence it was decided to run at lower speed and capture more frames in a single pattern\footnote{The length of the sequence is defined by the number of frames recorded in a single pattern.}.

\subsubsection{Experimental limitations}

\begin{figure}
	\centering
		\includegraphics[width=1.0\textwidth]{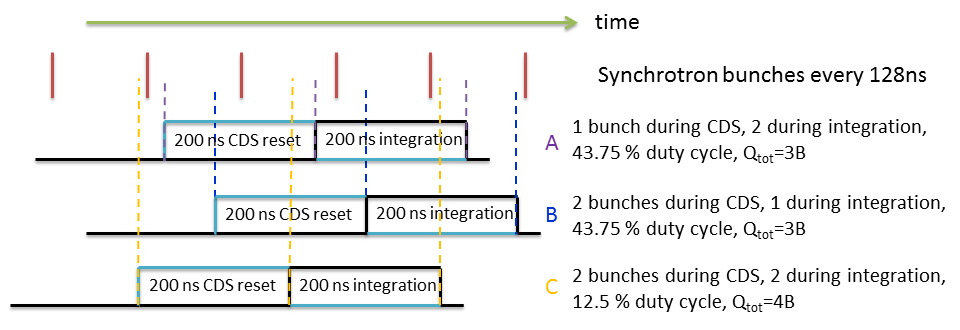} 
	\caption{The three different possibilities to distribute bunches every 128~ns during the CDS reset and integration phase.}
	\label{cycle}
\end{figure}

At the time of the test no bunch clock was available at P10. This led to the effects explained in the following paragraph.

In operation the preamplifier is sensitive to charges collected in the sensor material during the integration phase as well as during the Correlated Double Sampling (CDS) reset phase, which happens immediately before the integration phase. All charge (and equivalent noise charge, esp. the noise from the release of the preamp reset switch) that is accumulated during the CDS reset phase is removed by the CDS mechanism. This mechanism is ideally suited to reduce the effective noise of the system when properly synchronized to a pulsed source in such a way that no signal charge is accumulated during the CDS reset phase\footnote{When the operating conditions of the preamplifier change (i.e. by switching gain) during the integration time the double sampling is no longer correlated and its benefit is reduced. It should be noted that this has no impact on the detector performance as larger noise can be tolerated at the large signals required to trigger the gain switching.}.

During the experiment three possible situations, depicted in Figure \ref{cycle}, were possible: A) 1 bunch arrives during the CDS reset phase, 2 bunches during the integration phase, B) 2 bunches arrive during the CDS reset phase, 1 bunch during the integration phase or C) 2 bunches arrive both during CDS reset and integration phase.

Knowing that CDS reset and integration phase are each 200~ns long and that bunches come every 128~ns, one can calculate the duty cycles of case A) and B) to be 43.75\% each and case C) to be 12.5\%. If one bunch deposits, on average, a charge of $1B$, than cases A) and B) have the preamplifier swing from $0$ to $1B$ or $2B$ after the CDS reset phase and to $3B$ after the integration phase, while case C) swings the preamplifier from $0$ to $2B$ to $4B$. The output of the chip is $3B-1B=2B$ for case A), $3B-2B=1B$ for case B) and $4B-2B=2B$ for case C). It should be noted that the preamplifier in case C) operates at a different working point than in case A) which may lead to different ASIC outputs of cases A) and C), especially if the preamplifier leaves its linear regime for charges greater then $3B$.

\section{The P10 beamline}

\begin{figure}
	\centering
		\includegraphics[width=0.90\textwidth]{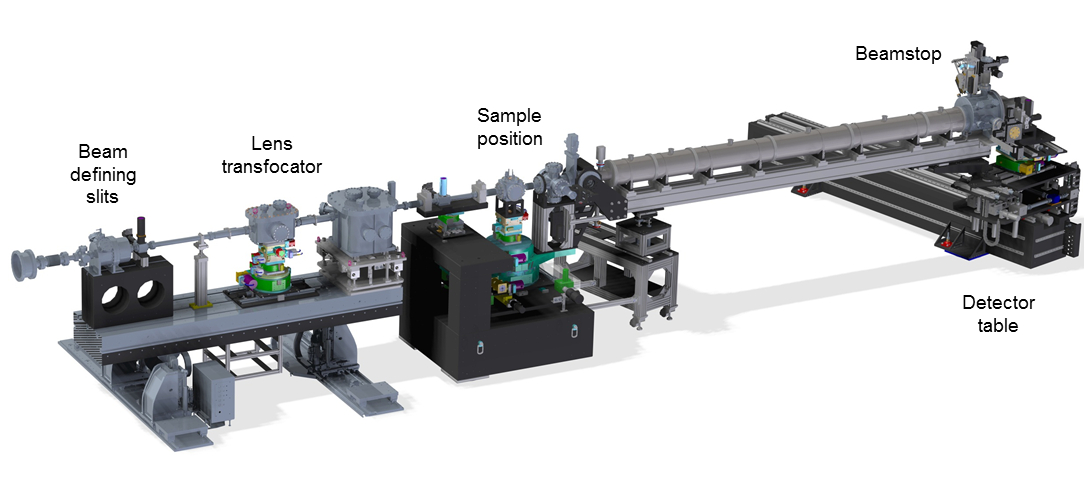} 
	\caption{Experimental setup at the P10 beamline.Upstream of the sketched setup the X-ray beam passes several optics elements including a monochromator and enters the sketched setup from the left side.}
	\label{setup}
\end{figure}

The coherence beamline P10 at PETRA III operates in the medium-hard x-ray regime (5-25~keV). For the AGIPD 0.4 chip test described here the photon energy was set to 7.05~keV and PETRA III was running in 60 bunch mode, resulting in a photon pulse approximately every 128~ns. The total flux at the sample position was estimated using a calibrated diode and a reference scatterer to be 1-2 $\times$ 10$^{11}$ photons/s.

The monochromatized beam was focused at sample location to a size of approximately 5~$\upmu$m $\times$ 3~$\upmu$m, which corresponds to a size of approximately 350~$\upmu$m $\times$ 175~$\upmu$m (vert. $\times$ horiz.) in the detector plane. A sketch of the experimental setup is shown in Figure \ref{setup}. The distance between the sample and the detector plane was 5.2~m.

The sample consisted of silica particles with a nominal radius of 250~nm suspended in water. The particle concentration was chosen such that the sample can be considered dilute (spheres are diffusing freely) and any effects of interparticle interactions can be neglected. 

\section{Results}


Except for the region very close to the primary beam, the images are virtually empty as the integration time of 200~ns is very short. Simply summing up frames would overemphasize the noise of the system, therefore the data was analyzed in a post-processing counting mode \cite{thresholding}.

In this counting mode every individual frame was analyzed separately. A global noise cut threshold was applied and the energy above threshold detected by each pixel was quantized to a discrete number of photons. This procedure is very similar to the signal processing inside the ASIC of photon counting detectors like the Pilatus \cite{pilatus} or the Medipix \cite{medipix}, with two notable exceptions: Multiple photon counts are registered correctly, and the threshold can be (re-)adjusted after the data is acquired. A detailed study of the impact of different threshold settings including the charge sharing effect has been presented in \cite{noise_paper}.



\subsection{Measurements without sample}

\begin{figure}
	\centering
	\begin{tabular}{@{}cc@{}} 
		\begin{tabular}{@{}c@{}} \includegraphics[width=0.30\textwidth]{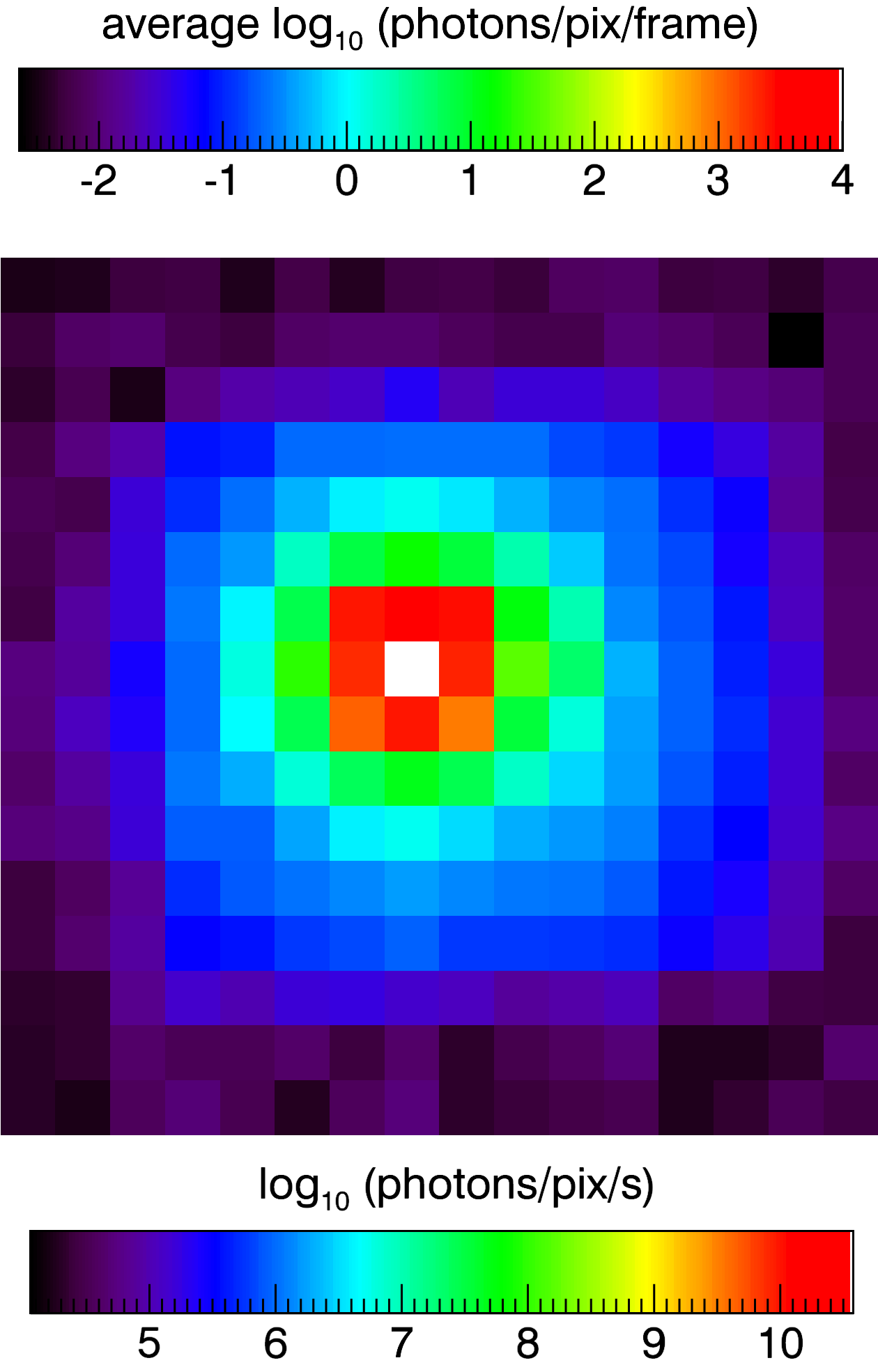}  \end{tabular} &
		\begin{tabular}{@{}c@{}} \includegraphics[width=0.65\textwidth]{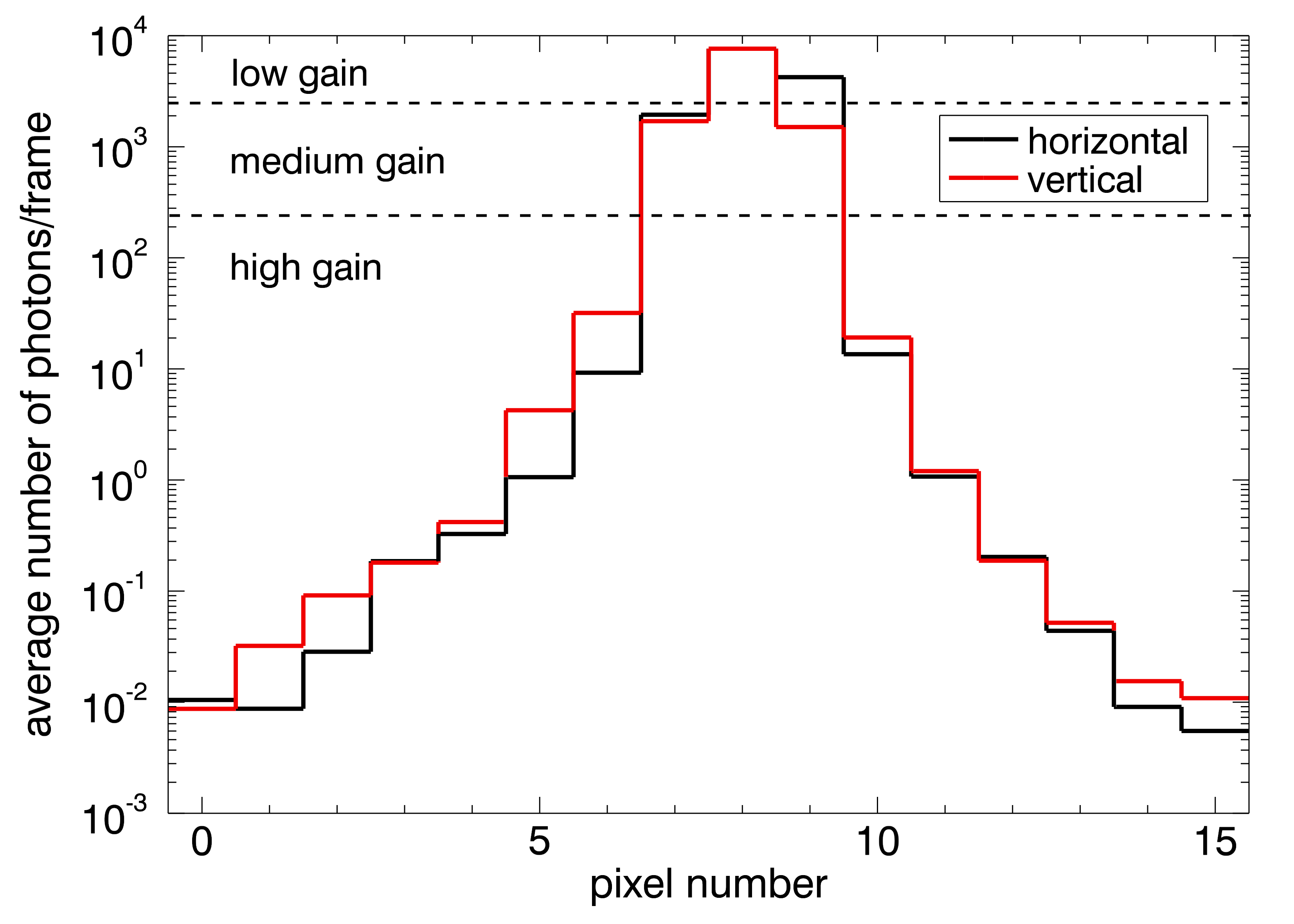}  \end{tabular} 
	\end{tabular}
	\caption{Left: Image of the direct beam. The color encodes the logarithm to the base 10 of the number of photons per pixel. Right: Line-cut of the left image along the horizontal and vertical direction through the most intense pixel. As indicated by the dashed lines, the image contains pixels in all three gain stages simultaneously. Both images show the data averaged from 3000 individual frames.}
	\label{fig1}
\end{figure}

\begin{figure}
	\centering
		\includegraphics[width=0.9\textwidth]{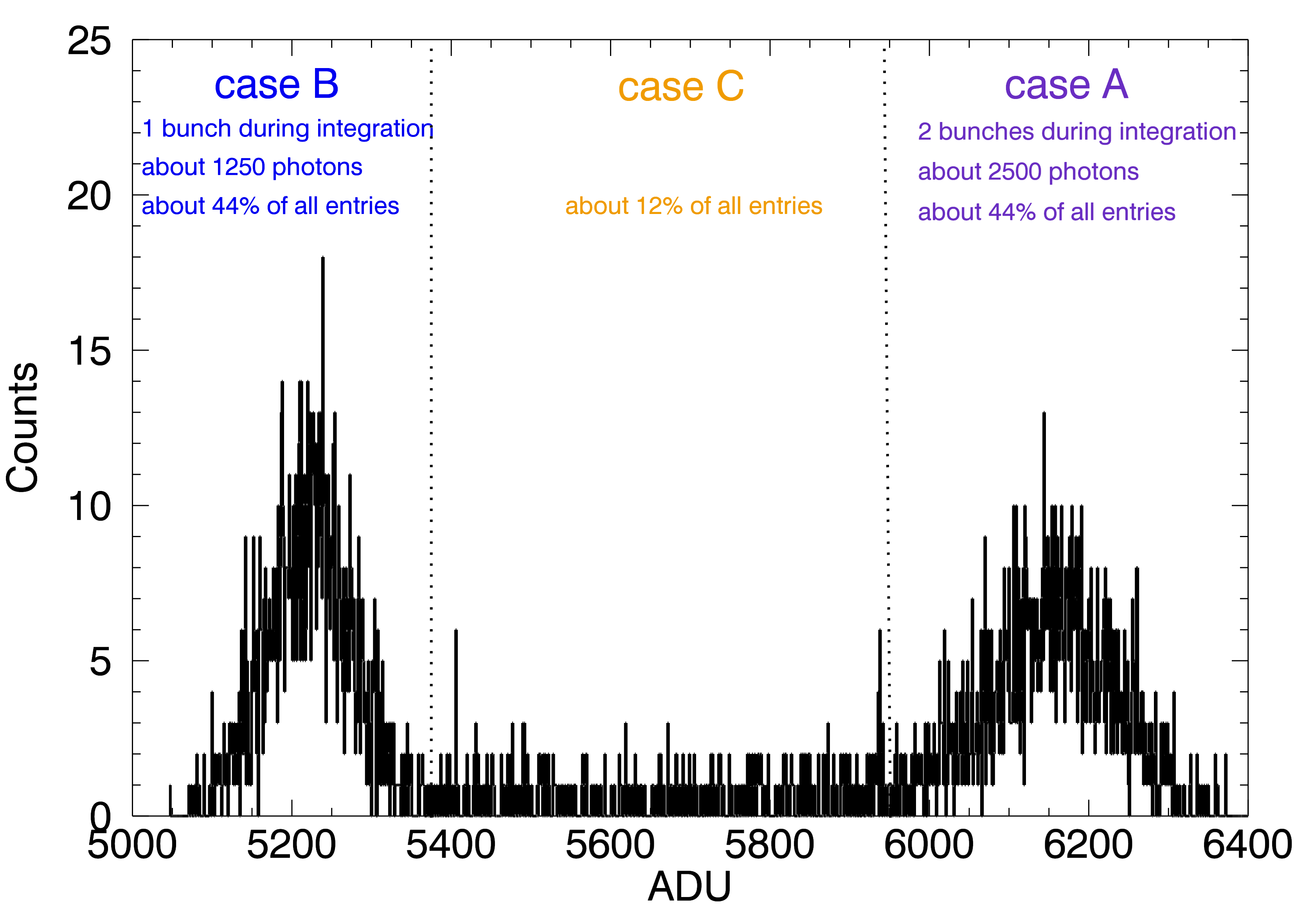} 
	\caption{Histogram of the analog output of the pixel up and left of the most intense pixel. The pixel is in medium gain mode. The two peaks indicate that either one or two bunches are impinging on the detector during the integration time. Having no photons in second gain would correspond to around 4300 ADU; note that when no photons are present the detector is in high gain mode, so this value is an extrapolation.}
	\label{histogram}
\end{figure}

In order to test the dynamic range, images of the direct beam were recorded. An image averaged from 3000 individual frames is displayed in Figure \ref{fig1}. The averaged image spans a dynamic range of more than 6 orders of magnitude. This is not obvious from individual frames, as the pixels far from the direct beam register no photons most of the time.

The X-ray beam was focused on the sample position, therefore it is defocused in the detector plane. The known slight asymmetry\footnote{Due to the different beam size and divergence in the horizontal and vertical plane.} of the primary beam is observed, as the beam is wider in vertical than in horizontal direction.

When looking at the histogram of the analog output of the pixel located one pixel to the left and up from the central pixel (shown in Figure \ref{histogram}), the pulsed nature of PETRA III can be seen clearly.  The pixel is in medium gain mode and shows the behavior explained previously. Two distinct peaks are easily observed and indicate that either one or two bunches are impinging on the detector during the integration time. The peaks contain about 44\% of the total number of counts each and the remaining 12\% of counts are distributed in between them. These percentages correspond to the duty cycle of the three different possibilities to distribute the bunches during the time the ASIC is sensitive to charge collection in the sensor, as depicted in Figure \ref{cycle}.

Imaging the direct beam is a test for the dynamic gain switching; in each individual image there are pixels in every gain stage. The central pixel is a special case, as it is operating outside of the linear regime. It is not completely saturated, as the distinct two peak nature of the histogram is still observable (not shown) and some corrections for the saturation behavior can be made.

Using the image displayed in Figure \ref{fig1} the total flux of the beam can be calculated to be approximately 0.5 $\times$ 10$^{11}$ photons/s. This is lower by a factor of approximately 3 than the estimate for the beam from the diode measurements, the reason for which is the partial saturation of the central pixel.

\subsection{Measurements with sample}

\begin{figure}
	\centering
	\begin{tabular}{@{}cc@{}} 
		\begin{tabular}{@{}c@{}} \includegraphics[width=0.35\textwidth]{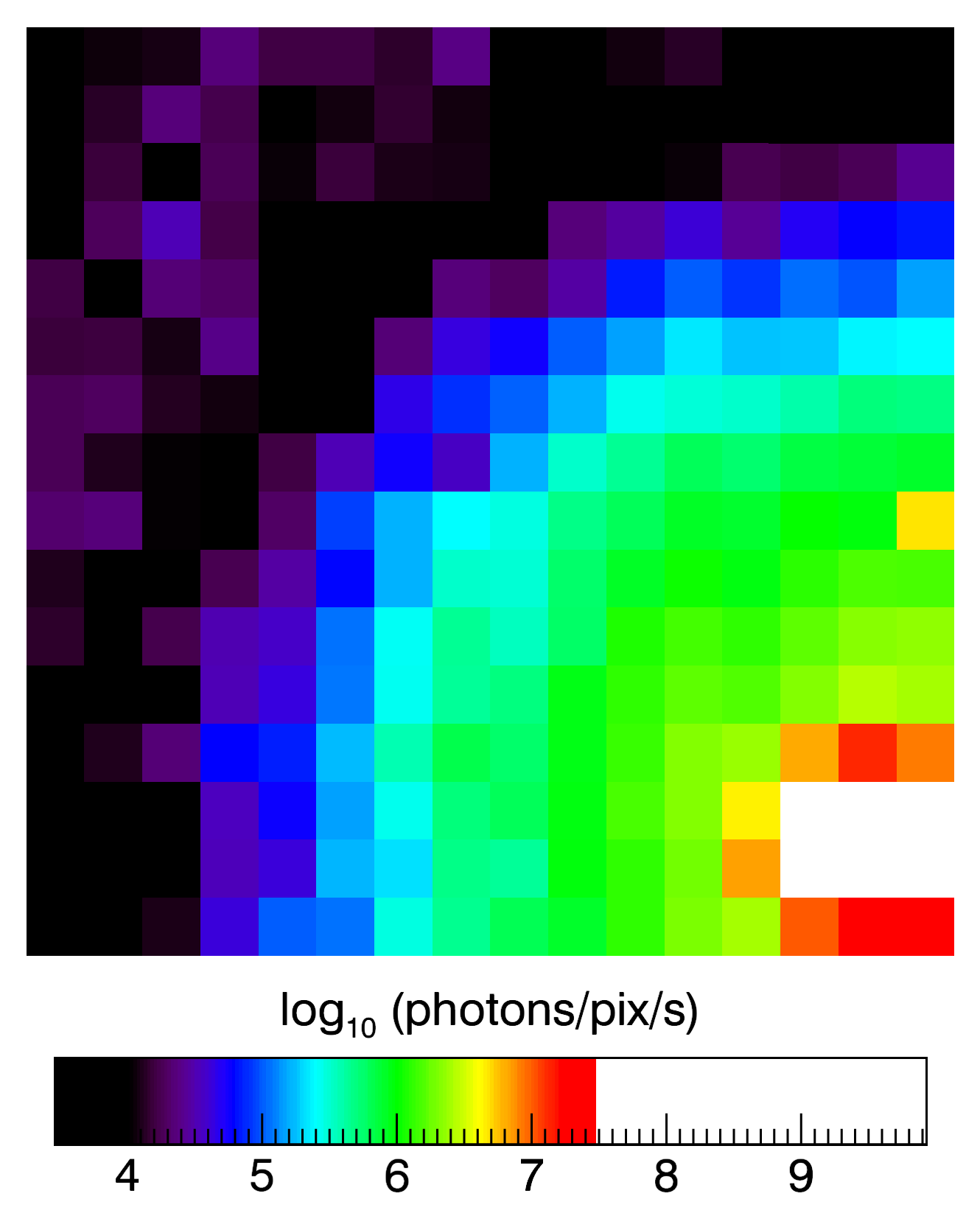}  \end{tabular} &
		\begin{tabular}{@{}c@{}} \includegraphics[width=0.60\textwidth]{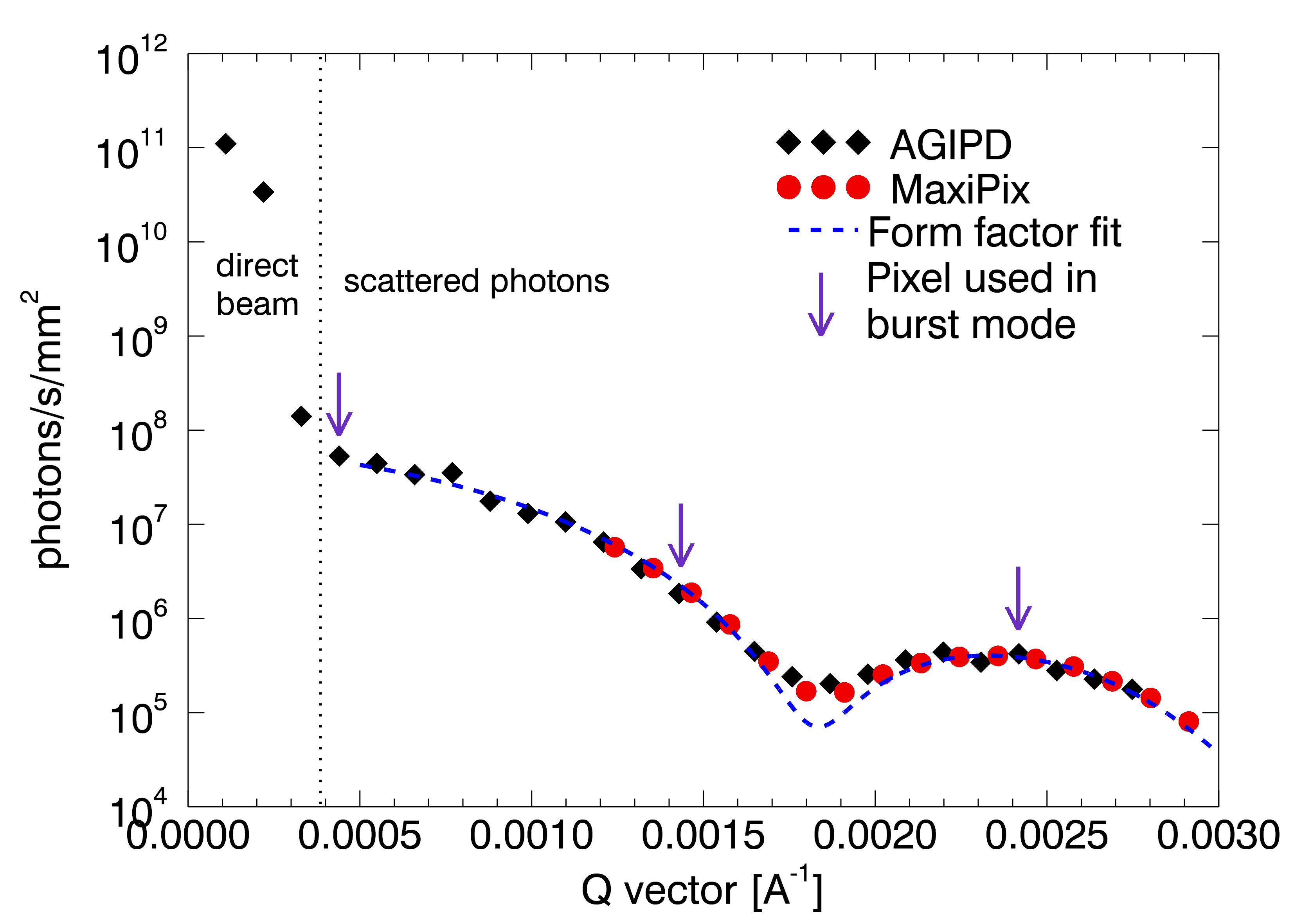}  \end{tabular} 
	\end{tabular}
	\caption{Left: Image of the scattering pattern of the sample. The color encodes the logarithm to the base 10 of the number of photons/s/pixel and has been restricted to the range from 4.0 to 7.5 to enhance the visibility of the ring structure. For this image 10000 individual frames were averaged. Right: Azimuthal average of the left image and comparison to data obtained with a MaxiPix detector. The dotted line indicates a particle form factor fit to the MaxiPix data.}
	\label{fig2}
\end{figure}

Studying colloidal particles, a common sample system for small angle X-ray scattering and XPCS experiments, is an efficient way of demonstrating the performance of the AGIPD 0.4 test chip. These particles can be synthesized from different materials in a large variety of sizes and shapes. Thereby their well known dynamics can be tailored to the specific task at hand \cite{westermaier}. In this experiment we will look at the particle form factor of the sample and at equilibrium fluctuations in the microsecond regime.

For the measurements a sample consisting of spherical colloidal particles with a nominal radius of 250~nm was inserted at the focus position of the beam.

The image displayed in Figure \ref{fig2} is an average of 10000 individual images and covers a dynamic range of more than 6 orders of magnitude. The color map of the image has been restricted to 3.5 orders of magnitude to enhance the visibility of the diffraction pattern with a local maximum at approximately 0.0022 \AA$^{-1}$. In addition, the same sample was imaged using a MaxiPix reference detector and the azimuthal average of the intensity distribution recorded with both detectors is shown in the right hand side graph of Figure \ref{fig2}. 

The dotted line is a fit of a spherical form factor (PQ-Fit) to the MaxiPix data. Only the MaxiPix data was used for this fit, as the recorded data extends to large scattering vectors due to the bigger detector area. The fit result for the particle radius is 245~nm.

It should be noted that no cross calibration between the two detector systems was performed. The AGIPD data reproduces the extrapolation of the PQ-Fit towards lower scattering angles and agrees with the intensity measured by the MaxiPix detector within the statistical errors.

In contrast to the MaxiPix detector, AGIPD 0.4 provides experimental data including the direct beam (three leftmost data points). Usually the beamline is operated with a beamstop to protect the detector from the direct beam, which blocks all signals below a certain scattering angle (approximately 0.001 \AA$^{-1}$  in this case).

\subsection{Burst mode}

\begin{figure}
	\centering
		\includegraphics[width=0.75\textwidth]{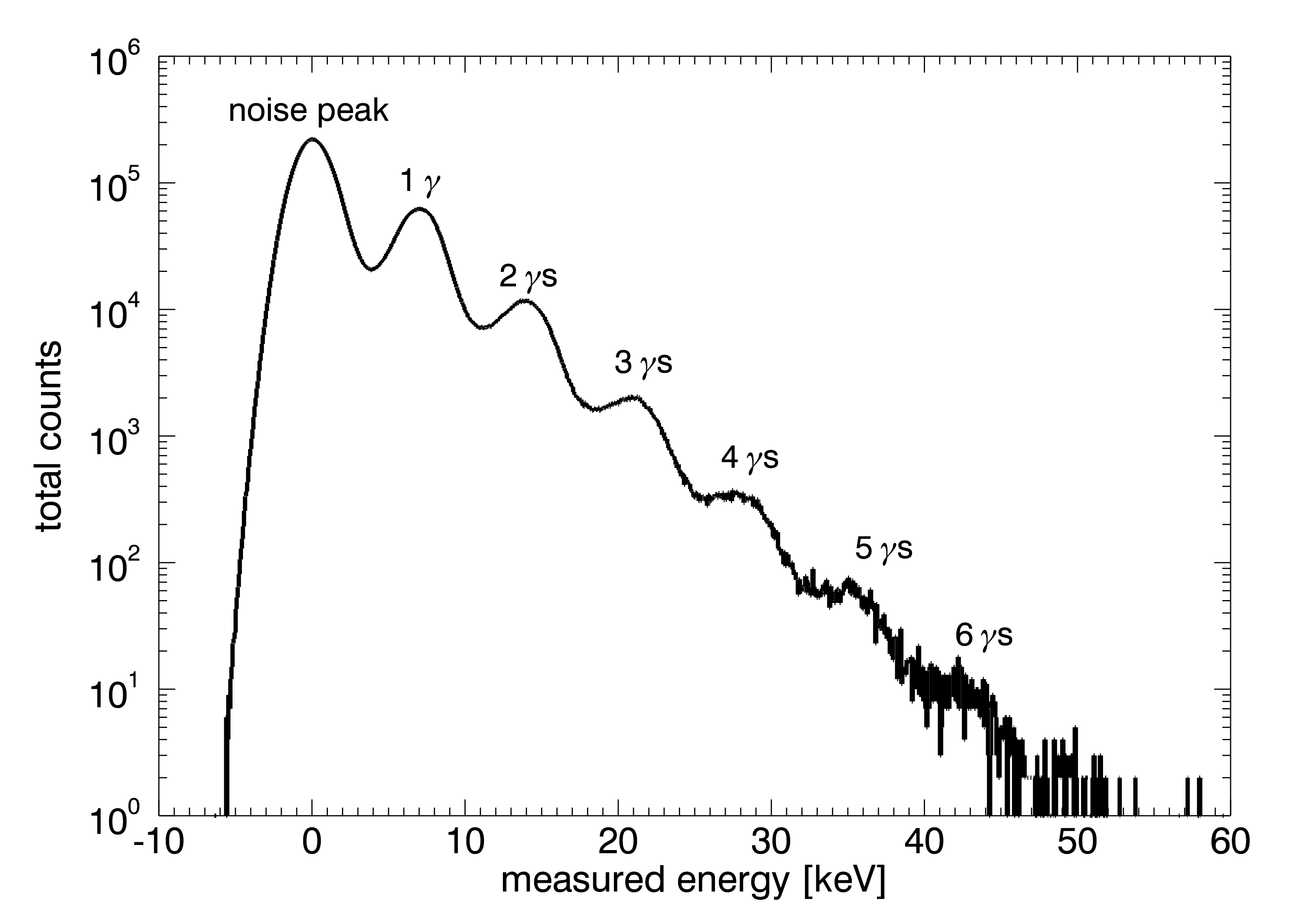} 
	\caption{Histogram of the largest data set taken with the XPCS pattern. The average intensity is 0.3 photons per 200~ns, corresponding to a count rate of 1.5 Mcps/pixel or 37.5 Mcps/mm$^2$. The individual peaks of photons of 7.05~keV are clearly distinguishable. The rms noise of this pixel is 320 electrons.}
	\label{single}
\end{figure}


Three pixels, at 0.00044 \AA$^{-1}$, 0.0014 \AA$^{-1}$ and 0.0024 \AA$^{-1}$, were selected to perform XPCS experiments using the dedicated XPCS pattern. The pixels were measured individually, one after the other.

Figure \ref{single} shows the counting statistics for the pixel at 0.00044 \AA$^{-1}$. Charge integration during the CDS reset phase is negligible, as the average count rate of 0.3 photons/frame\footnote{Due to the short integration time of 200~ns this corresponds to 1.5 Mcps/pixel, or 37.5 Mcps/mm$^2$.} uses only a very tiny fraction of the dynamic range of the first gain stage (extending to about 1.1~MeV deposited energy). 

The investigated pixel has a noise of approximately 320 electrons rms, which is consistent with previous noise measurements \cite{noise}. This pixel fulfills the criteria for single photon detection at the European XFEL \cite{noise_paper}. 
Figure \ref{single} shows a clear separation up to 6 photons, at which point statistics limit the unanimous identification of further peaks. 


\subsubsection{Intensity autocorrelation}

The intensity autocorrelation function ($g_2$ function) is both a measure of the correlation time and the speckle contrast of a system. Details on this analysis type can be found in textbooks like \cite{xpcs_scm_book}, specific details for XPCS using the AGIPD can be found here \cite{XPCS1, XPCS2}.

With our setup the bunch frequency of PETRA III cannot be measured directly; however, the pulsed nature is still observable with a lower apparent frequency that is determined by the aliasing effect.

The apparent frequency $f_{alias}$ expected from the aliasing effect can be calculated from the true frequency $f_{true}$, the sampling frequency $f_{sampling}$ and an integer number $N$:

\begin{figure}
	\centering
		\includegraphics[width=0.75\textwidth]{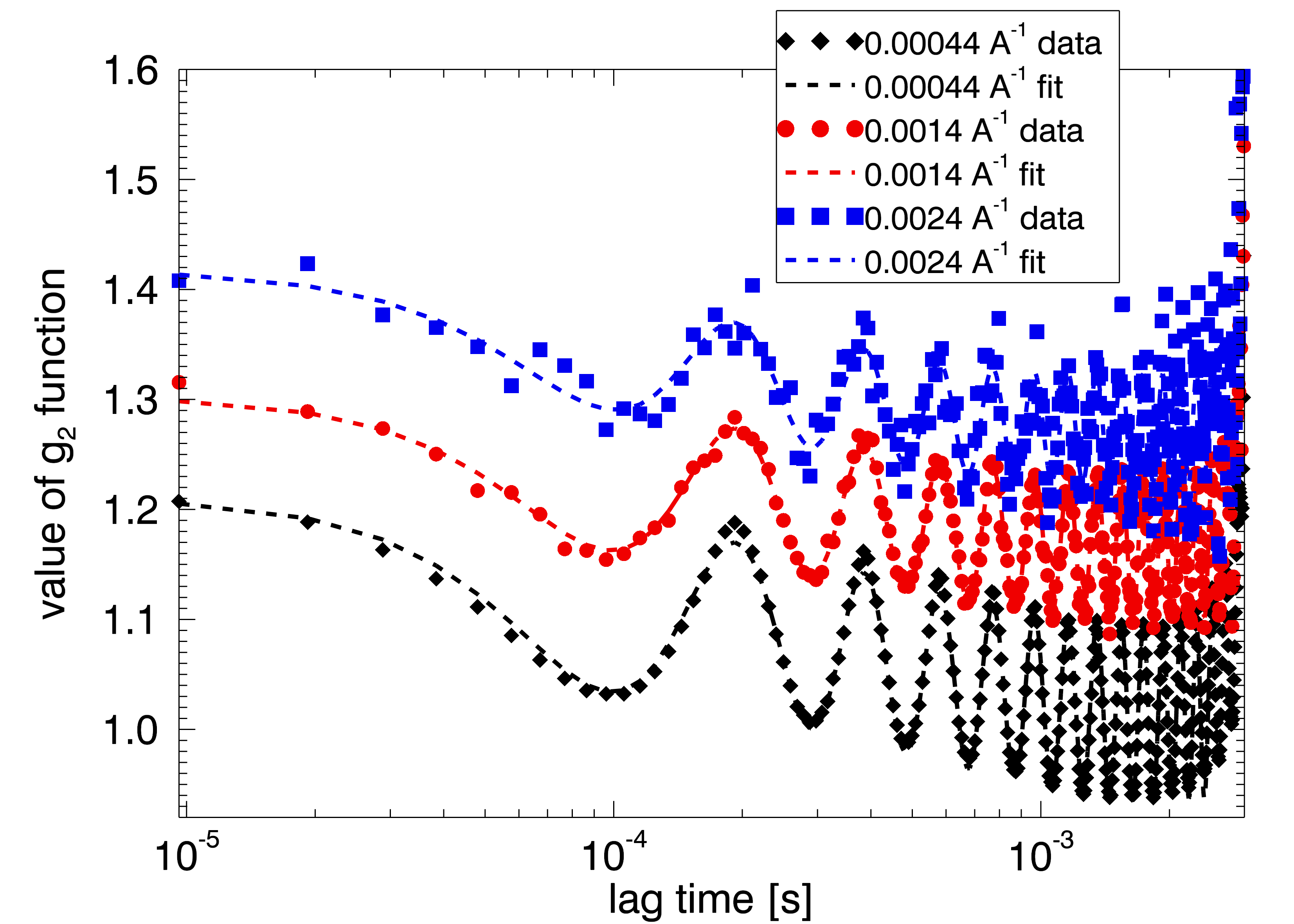} 
	\caption{$g_2$ functions derived from the XPCS data. The different scattering angles are indicated by the color. Symbols indicate the data points from the measurement, lines analytic fits to the data. The $g_2$ functions have been horizontally offset for better visibility. AGIPD 0.4 allows determination of correlations on the microsecond timescale. The observed oscillations originate in the aliased frequency of the storage ring and the pixel readout rate.}
	\label{g2}
\end{figure}

\begin{equation}
	f_{alias} = \left| f_{true} - N f_{sampling} \right |. \label{alias1}
\end{equation}

Using the orbital length\footnote{This is the 2.3~km circumference of PETRA III. A high precision value of this dimension (accurate to 1~cm) was calculated from the radio frequency of the machine.} of the storage ring $L_{PETRA}$, the speed of light c, $M= 60$ the number of bunches, the sampling frequency $ f_{sampling}= 1/9.602\mu s = 104.145 kHz$ and $N = 75$ we obtain the following frequency expected for our experiments:

\begin{equation}
	f_{alias} = \left| \frac{cM}{L_{Petra}} - N f_{sampling} \right | = 5.185 \; kHz. \label{alias2}
\end{equation}

%
%

Figure \ref{g2} shows the $g_2$ functions derived from the data (symbols) accompanied by lines representing the analytic fit to the data points. 

It is readily observable that the $g_2$ functions do not follow the expected monotonic decrease, but show a characteristic oscillation with a local minimum at a lag time of approximately 10$^{-4}$~s. The analytic fits identify the oscillation frequency to be 5.18~kHz, which is the expected aliased bunch repetition frequency. 

Additionally, all functions show a 'hockey stick' behavior for lag times above approximately 10$^{-4}$~s. They decrease more than expected from the sample dynamics\footnote{This is only visible when compared to detailed simulations, which are beyond the scope of this paper.}, show a local minimum (overlaid by the oscillations) between 1 and 2 $\times$ 10$^{-3}$~s and increase to an off-scale value for the last few data points. This behavior is caused by the limited number of frames.



Emphasis should be put on the timescale of the study, not on the studied physics. The relevant timescales that are within reach of the correlation analysis are on the order of several microseconds. In addition it should be noted that for this type of analysis the reliable detection of individual photons is mandatory and achieved by AGIPD 0.4.

\section{Observations on radiation damage}
As operation without a beam stop is very damaging to many detectors, the detector was carefully checked for radiation damage effects.

Most notable is the increase of the sensor leakage current. The sensor was not optimized for radiation hardness, therefore the current increase is readily observed. However due to the short integration time the increased leakage current is completely negligible, the noise of the system is dominated by other sources \cite{noise}, even after irradiation. 

Several ASIC parameters were carefully monitored for changes during the experiments, but none were observed. This is in accordance with results from dedicated irradiation campaigns \cite{marras}, which do not show any functional deterioration of the chip (just a slight noise increase) for doses up to 10~MGy. Taking the attenuation of 7~keV photons in the sensor material into account, it would have taken about 2 weeks of non stop irradiation to reach this dose in the most intensely illuminated pixel.

\section{Summary}

An AGIPD 0.4 test chip was installed at the P10 beamline and several prototypical experiments were performed. Although it is a prototype, all components relevant for operation at the European XFEL are present.

The direct beam was imaged successfully, thereby demonstrating that the large dynamic range required for XFEL applications (single photons and 10$^4$ 12~keV photons simultaneously in a single frame) is achieved.

The single photon sensitivity required for experiments at the European XFEL (ENC of less than 340 electrons) was shown and individual 7~keV photons could be distinguished. Please note that in the original design of the detector operation at a fixed beam energy of 12.4~keV was anticipated.

The intensity fluctuation analysis of the colloidal sample demonstrated the burst mode capability, which is essential for usage at the European XFEL. Correlations at the level of several microseconds were observed, which makes AGIPD 0.4 one of the fastest 2D imaging detectors currently available.



\acknowledgments

The authors would like to thank DESY and especially the PETRA III facility for access to the beamline, the European XFEL for co-funding the development of AGIPD, the entire team that designed, produced and assembled AGIPD 0.4 and Bernd Struth and Michael Sprung for their input.

\end{document}